\documentclass[amsmath,trackchanges, twocolumn]{aastex702}

\usepackage{booktabs}
\usepackage{tabularx}

\usepackage{tikz}
\usepackage{multirow}
\usepackage{makecell}
\usepackage{amsmath}
\usepackage{siunitx}
\usepackage{booktabs}
\usepackage{afterpage}
\usepackage{xcolor}

\usepackage{array}

\usepackage{array}
\newcolumntype{L}[1]{%
  >{\raggedright\arraybackslash\binoppenalty=9999\relpenalty=9999}p{#1}}

\newcolumntype{L}[1]{>{\raggedright\arraybackslash}p{#1}}
\newcolumntype{C}[1]{>{\centering\arraybackslash}p{#1}}
\newcolumntype{R}[1]{>{\raggedleft\arraybackslash}p{#1}}
\usetikzlibrary{arrows.meta, positioning, fit, calc, shapes.geometric}

\begin{document}

\title{Asteroids Impacting the Solar System Planets and the Moon. III. Real Impacts from Known Objects and LSST Discovery Predictions}

\author[0009-0002-7847-8082]{Qifeng Cheng}
    \affiliation{Department of Physics, Duke University, Durham, NC 27708, USA}
    \email{qifeng.cheng@duke.edu}  
\author[0000-0002-4934-5849]{Daniel Scolnic}
      \affiliation{Department of Physics, Duke University, Durham, NC 27708, USA}
      \affiliation{Department of Electrical and Computer Engineering, Duke University, Durham, NC 27708, USA}
      \email{dan.scolnic@duke.edu}
\correspondingauthor{Qifeng Cheng,
\href{mailto:qifeng.cheng@duke.edu}{qifeng.cheng@duke.edu}}

\begin{abstract}

Impact rates predicted using pre-impact source-population models and those inferred from observations disagree by a factor of a few to $\sim200$ across the Solar System planets and the Moon. An exciting opportunity for an additional observational constraint is the discovery of asteroids before impact and their tracking until impact. We evaluate this prospect based on both known objects and forecasts of the Vera~C.\ Rubin Observatory Legacy Survey of Space and Time (LSST). First, we identify impacts from known objects, confirming multiple past cases and predicting two coming impacts: \texttt{2015~FK488}, a $\sim1.8$~km Centaur whose nominal orbit predicts an impact with Jupiter in $\sim200$~yr, and \texttt{2022~KG1}, a $<10$ m NEO recently removed from the impact-risk lists but whose nominal trajectory leads to an Earth impact in $\sim260$~yr. Both orbits are poorly constrained. Second, assuming the model-based impact rate, we predict that LSST will discover 7 of the 59 synthetic impactors quantified in our previous analysis. All of the discovered impactors will impact Jupiter, and they have a size range of $\sim200$--$2000$ m and expected discovery-to-collision lead times of $1.94-257.4$ yr. Two of them are discovered inside Jupiter's Hill sphere, offering rare opportunities to observe the transition from a temporary bound orbit to impact before it happens. If instead we assume the observation-inferred impact rates, LSST will at most discover $\sim20$ future Earth impactors ($>10$ m) and $\sim300$ future Jupiter impactors ($>10$ m), with a wider range on both ends for discovery-to-collision lead times. These results demonstrate that LSST can identify a subset of planetary impactors years to centuries before collision, providing a new observational pathway for testing source-population models and the impact-rate mismatch.

\end{abstract}

\section{Introduction}

The rate at which asteroids or comets strike the Solar System planets and the Moon is constrained by two lines of evidence, and they do not agree. Dynamical models propagate synthetic survey-debiased populations and predict collision rates, while direct observations record impact events through bolides, flashes, seismic signals, and fresh craters. In Papers I and II we compared the two on a common size threshold and found that they differ by a factor of a few to two hundred across Earth, the Moon, Mars, and Jupiter. 

Between these two approaches, objects that are already known and tracked but have not yet impacted provide an additional sample for constraining impact physics and, for the purpose of this work, impact rates. One of the best examples of pre-discovered impactors, and the only case involving a planetary body other than Earth, is Comet Shoemaker-Levy 9 (hereafter SL9). It was discovered in March 1993 and recognized within two months to be on a collision course with Jupiter \citep{Marsden1993, ChodasYeomans1996}. Its fragments struck the planet in July 1994 after more than a year of advance warning, making it the only kilometer-scale body whose impact was both predicted and observed \citep{Hammel1995, ChodasYeomans1996}. It constrains the modeled Jovian impact flux both directly, through the SL9-class event frequency derived from the fragment-breakup physics \citep{RoulstonAhrens1997}, and as a benchmark on the ecliptic-comet impact rate at kilometer sizes \citep{Levison2000}. It also marks the large-size observation anchor for impact-rate derivations \citep{hueso2018small}. Its early discovery also enabled coordinated observing campaigns that captured the impacts in detail \citep{Carusi1994,ChodasYeomans1996}, providing opportunities to constrain energy deposition and luminous efficiency of an impact on which the flash-based impact rates depend. 


On Earth, the known examples involve smaller objects discovered with much shorter warning times. As of August 2026, twelve asteroids had been discovered prior to Earth impact, all with sizes $<10$ m and warning times ranging from $\sim$2 to $\sim$20 hours. Early discovery yields a telescopic absolute magnitude that pairs with an independently measured impact energy from satellite sensors, and the two measurements together calibrate the conversion between the survey-inferred population and the observed impact flux \citep{Chow_2026}. The follow-up observations for objects that surveys successfully track before impact also enable compositional characterization, such as the meteorite- or fireball-derived albedos of 2008~TC$_3$ \citep{Jenniskens2009} and 2022~WJ$_1$ \citep{Kareta2024}. These measurements help calibrate the conversion from absolute magnitude $H$ to diameter, thereby reducing a key source of uncertainty in inferred size-dependent impact rates.




The first goal of this paper is therefore to run real asteroids and comets through the Paper I impact-search method and determine how many impacts it predicts. We draw these objects from JPL's Small-Body Database (SBDB\footnote{\url{https://ssd.jpl.nasa.gov/tools/sbdb_query.html}};\citealt{Giorgini2015}), the most comprehensive catalog of discovered small bodies. SBDB provides computed orbits from the full astrometric record published by the Minor Planet Center, including radar ranging where available, using debiased, weighted observations. 

Meanwhile, as upcoming surveys expand the known populations, pre-impact observations could provide increasingly important constraints on planetary impact rates. LSST \citep{LSST} will drive that growth, increasing the number of known small bodies by a factor of $4-9$ \citep{Kurlander2025}. The second goal of this paper is therefore to predict how many intrinsic impactors LSST will discover. We make this prediction using the dedicated survey simulator \texttt{Sorcha}, which models its Solar System survey completeness under realistic observing cadences, magnitude limits, trailing losses, and moving-object discovery criteria \citep{merritt2025sorcha}. We use \texttt{Sorcha} to forward model the synthetic close-encounter and impactor populations from Paper I. The recent LSST Data Preview 2 (DP2) additionally makes it possible to test any predicted discoveries against real detections.

We structure our paper as follows. In Section~\ref{sec:known}, we describe our query of known objects and identify impactors. In Section~\ref{sec:lsst}, we describe our LSST simulation setup and report the predicted discoveries. We discuss the unique pre-impact observing opportunities offered by LSST and the limitations of unmodeled cometary effects in Section~\ref{sec:discussion}. 






\section{Impactors from Known Small Bodies}
\label{sec:known}

\subsection{JPL SBDB Catalog Query and Impact Identifier Setup}

We retrieve the orbital and physical parameters of the known small body populations from the JPL SBDB Query API \footnote{\url{https://ssd-api.jpl.nasa.gov/sbdb_query.api}} for the same five source populations used in Paper I, including near-Earth Objects (NEOs), main-belt asteroids (MBAs), Jupiter-family comets (JFCs), Centaurs, and trans-Neptunian objects (TNOs). Number breakdowns are shown in Table~\ref{tab:sbdb_population}. 

We query each population by sub-classes of orbit-class codes and boundaries defined by SBDB. The NEO population comprises four subclasses defined by perihelion and aphelion distances: Atiras (IEO), with orbits entirely within Earth's orbit ($Q<0.983$ au); Atens (ATE), with $a < 1$~au and Earth-crossing aphelia; Apollos (APO), with $a > 1$~au and Earth-crossing perihelia; and Amors (AMO), with perihelia just exterior to Earth's orbit ($1.017 < q < 1.3$~au). The MBA population comprises three subclasses set by semi-major axis: the inner (IMB, $a < 2.0$~au), main (MBA, $2.0 < a < 3.2$~au), and outer (OMB, $3.2 < a < 4.6$~au) belt. The JFC population comprises a dynamical subclass (JFc), defined by the Jupiter Tisserand parameter $2 < T_{\mathrm{J}} < 3$, and a classical subclass (JFC), defined by orbital period $P < 20$~yr. The Centaur (CEN, $5.5 < a < 30.1$~au) and TNO ($a > 30.1$~au) populations each correspond to a single orbit class.

SBDB does not provide a dynamical subclass for the scattering disk objects, so we flag scattering candidates among the TNOs by 
\begin{equation}
    a > 50~\mathrm{au} \quad \mathrm{and} \quad q < 40~\mathrm{au},
    \label{eq:scattering_filter}
\end{equation}
which selects orbits with large semi-major axes but perihelia near Neptune. We flag $1{,}665$ scattering TNOs of the $7{,}288$ TNOs ($22.8\%$). We emphasize that Equation~\ref{eq:scattering_filter} is an approximation, and the dynamical classification, such as one following \cite{gladman2008}, would require resonance identification or orbit integration, which is beyond the scope of this work. 

We retrieve albedos and sizes from SBDB, and cross-check against the published (Near-Earth Object) Wide-field Infrared Survey Explorer \citep[WISE/NEOWISE;][]{wright2010, mainzer2011_neowise} catalogs from the VizieR service \footnote{\url{https://vizier.cds.unistra.fr}}. For objects with an unreported diameter, we derive one from Equation~1 of Paper I using each object's absolute magnitude and albedo. Absolute magnitudes are almost entirely measured ($1948:1$)\footnote{SBDB reports comet total magnitude as $M_1/M_2$ rather than $H$, so absolute magnitudes are unavailable for the JFC population}. We adopt measured albedos where available and derive albedos otherwise by following the modeled-albedo method in Paper I. The measured-to-assumed ratio of albedo is $1:10$. This procedure yields diameters for $90.9\%$ of the population, raising total coverage to $99.6\%$ when combined with the measured diameters.

We adopt the object-level periapsis-distance-based impact estimation procedure in Paper I. We propagate the known objects through 2325 using \texttt{ASSIST}/\texttt{REBOUND}, identify close encounters entering the Hill spheres of the eight planets and the Moon, calculate periapsis distances at hour-scale timesteps, and flag objects whose periapsis lies within the target body's physical radius, taken at the 1-bar atmospheric pressure-level for the giant planets. The thresholds and parameters adopted at each step are detailed in Paper I.


\begin{table}
\centering
\caption{Number of known small bodies queried from the JPL Small-Body Database (SBDB)}
\label{tab:sbdb_population}

\begin{tabular}{@{}llr@{}}
\hline
Population & SBDB subclass & Number of objects \\
\hline
NEO & IEO (Atira)  & 38 \\
    & ATE (Aten)   & 3,432 \\
    & APO (Apollo) & 23,973 \\
    & AMO (Amor)   & 14,709 \\
    & \textbf{Subtotal} & \textbf{42,152} \\
\hline
MBA & IMB (Inner main belt) & 32,841 \\
    & MBA (Main belt)       & 1,376,481 \\
    & OMB (Outer main belt) & 50,062 \\
    & \textbf{Subtotal}     & \textbf{1,459,384} \\
\hline
JFC & JFc (dynamical) & 829 \\
    & JFC (classical) & 17 \\
    & \textbf{Subtotal} & \textbf{846} \\
\hline
Centaur & CEN & 1,046 \\
TNO     & TNO & 7,288 \\
\hline
\multicolumn{2}{l}{\textbf{Total}} & \textbf{1,510,716} \\
\hline
\end{tabular}
\tablecomments{Data were queried on August 15th, 2026. The scattering-TNO subset is isolated from the full TNO population using the orbital filter in Equation~\ref{eq:scattering_filter}.}
\end{table}

\subsection{Identified Impactors}

We discover 15 impact events, of which 13 are confirmed. They include twelve single-body Earth impactors and one multi-fragment SL9 collision with Jupiter. The most recent case is \texttt{2026~JN4}, with a confirmed atmospheric entry over the Arafura Sea on May 15, 2026, and our predicted impact time falls within 10.3 minutes of the observed impact time, with an estimated diameter of $\sim0.75$~m ($H=33.38$). The remaining 11 are confirmed NEO impactors discovered between 2008 and 2025, namely, 2008 TC3, 2014 AA, 2018 LA, 2019 MO, 2022 EB5, 2022 WJ1, 2023 CX1, 2024 BX1, 2024 RW1, 2024 UQ, 2024 XA1, and our predicted impact times differ from the observed times by $\sim8$ minutes on average. The SL9 event comprises 21 JFC fragment impacts on Jupiter for which orbit solutions exist. Nineteen match the published 1994 impact times to within $\sim20$--$26$~minutes. Two are exceptions. The predicted impact time for fragment~A differs substantially from the reported time ($>100$ years, likely an anomaly), while we cannot provide the time difference for fragment~P1 because its observational tracking was lost before reaching Jupiter's surface. For the confirmed impact cases, the predictions generally agree well with observations because the observational and impact epochs are separated by only a short integration interval ($\sim17$ hours on average for the 12 Earth impactors and $~72.9$ days on average for the SL9 fragments).

We predict two impacts, but both are too poorly constrained and cannot presently be verified. \texttt{2022~KG1}, predicted to impact Earth in $\sim260$ years, was removed from the JPL Sentry risk list in 2022, and its 7-day arc and condition-code-5 orbit do not support a precise multi-century extrapolation. The single Centaur impactor, \texttt{2015~FK488}, is a $\sim1.8$~km object predicted to strike Jupiter in 2227, and is the only predicted impactor above 10~m. However, we are unable to verify this case because we currently lack a Jupiter-impact monitoring system and its orbit is poorly constrained by 13 observations over a 7-day arc, the same limitation as \texttt{2022~KG1}. Both predictions require follow-up observations and full impact-probability propagation before they can be assessed. 

\subsection{Objects Within One Lunar Distance}

Across the full integration to the year 2325, the periapsis method flags 564 NEO–Earth encounters within one lunar distance (LD; $384{,}400$~km). Of these, 412 (73\%) fall within JPL's Close-Approach Data (CAD). However, CAD does not publish sub-LD predictions beyond the year 2193, leaving the remaining 152 (27\%) unverifiable regardless of their intrinsic accuracy. Within the 412-object verifiable window, 243 (59\%) agree with the published sub-LD classification and 169 (41\%) do not.

Two possible effects may contribute to the 41\% misclassifications. First, the orbits of a substantial fraction of objects are poorly constrained (condition codes 7–9, observational arcs of days) to compute reliable predictions at this lead time, even inside the 2193 search window. Second, our pipeline propagates the nominal orbit without covariance, so agreement degrades with longer lead time. Near-term encounters (through 2035) match well (mean offset $\sim36$~min, maximum $2.4$~d across 173 objects), whereas century-scale encounters diverge as small initial uncertainties amplify through repeated Earth and Venus flybys. 

\section{LSST Simulations}
\label{sec:lsst}
\subsection{Simulation Method and Setup}

To incorporate observational selection effects and predict LSST discoveries based on model-based impact rates, we forward model the simulated planetary close-encounter and impactor populations from Paper I with Sorcha \citep{merritt2025sorcha}, and analyze discovered and missed objects. 

Given an input synthetic population, \texttt{Sorcha} propagates each object through the simulated survey cadence and determines which objects fall within LSST pointings and are bright enough to be detected.  This procedure accounts for telescope pointing data, detection limits, photometric uncertainty, and trailing losses of moving objects \citep{merritt2025sorcha,holman2025sorcha}. We use the updated survey cadence \texttt{baseline\_v5.1} by LSST Operations Simulator (OpSim) \citep{delgado2014opsim}, and adopt the same simulation setup as \citet{Cheng_2026}, including a median single-visit depth of $m_r \simeq 23.95$, a per-visit exposure time of 30~s, the standard internal randomization and brightness filtering, and the default linking criteria of at least three tracklets \citep{VeresChesley2017b} within 15 days \citep{Ivezic2019}. Objects passing all these criteria are classified as LSST-discovered. We classify missed objects into three categories: never within an LSST pointing, too faint to detect, or detected but not successfully linked \citep{Cheng_2026}.

\begingroup
\sisetup{group-separator={,}, group-minimum-digits=4}
\begin{deluxetable*}{llrrrrr|rrrrr}
\tabletypesize{\scriptsize}
\tablewidth{0pt}
\setlength{\tabcolsep}{4pt}
\renewcommand{\arraystretch}{1.4}

\tablecaption{LSST detection for close encounters and impactors by source population and planet.\label{tab:loss_modes}}

\tablehead{
\colhead{} &
\colhead{} &
\multicolumn{5}{c|}{Close encounters} &
\multicolumn{5}{c}{Impactors} \\
\colhead{Group} &
\colhead{Planet} &
\colhead{$N_\mathrm{cross}$} &
\colhead{$N_\mathrm{ptg}$} &
\colhead{$N_\mathrm{det}$} &
\colhead{$N_\mathrm{disc}$} &
\multicolumn{1}{c|}{$\epsilon_\mathrm{disc}$} &
\colhead{$N_\mathrm{imp}$} &
\colhead{$N_\mathrm{ptg}$} &
\colhead{$N_\mathrm{det}$} &
\colhead{$N_\mathrm{disc}$} &
\colhead{$\epsilon_\mathrm{disc}$}
}
\startdata
\shortstack[l]{NEOs}
  & Mercury & 387          & 294 & 29          & 12         & 3.10\% & 0 & \nodata & \nodata & \nodata & \nodata \\
  & Venus   & \num{42485}  & \num{26419} & \num{3369}  & 759        & 1.79\% & 6       & 4 & 0 & 0 & 0.00\% \\
  & Earth   & \num{120093} & \num{75368} & \num{18297} & \num{4910} & 4.09\% & 5       & 3 & 1 & 0 & 0.00\% \\
  & Mars    & \num{34006}  & \num{21277} & \num{3182}  & 784        & 2.31\% & 0 & \nodata & \nodata & \nodata & \nodata \\
  & Jupiter & \num{30062}  & \num{18837} & \num{1080}  & 346        & 1.15\% & 11      & 9 & 0 & 0 & 0.00\% \\
  & Saturn  & 8            & 6 & 0           & 0          & 0.00\% & 0 & \nodata & \nodata & \nodata & \nodata \\
\tableline
\shortstack[l]{MBAs\tablenotemark{a}}
  & Mars    & 597    & 596 & 240    & 117   & 19.60\% & 0 & \nodata & \nodata & \nodata & \nodata \\
  & Jupiter & 8,917  & 8,890 & 5,442  & 2,798 & 31.38\% & 19      & 19 & 12 & 7 & 36.84\% \\
  & Saturn  & 239    & 238 & 128    & 60    & 25.10\% & 0 & \nodata & \nodata & \nodata & \nodata \\
\tableline
\shortstack[l]{JFCs}
  & Jupiter & \num{1398214} & \num{1395852} & \num{3554} & 788 & 0.056\% & 17 & 17 & 0 & 0 & 0.00\% \\
  & Saturn  & \num{78782}   & \num{78513}   & 90         & 20  & 0.025\% & 1  & 1  & 0 & 0 & 0.00\% \\
  & Earth   & \num{3508}    & \num{3486}    & 85         & 18  & 0.513\% & 0  & \nodata & \nodata & \nodata & \nodata \\
  & Mars    & \num{1374}    & \num{1373}    & 7          & 1   & 0.073\% & 0  & \nodata & \nodata & \nodata & \nodata \\
  & Venus   & 807           & 798           & 11         & 3   & 0.372\% & 0  & \nodata & \nodata & \nodata & \nodata \\
\tableline
\shortstack[l]{Centaurs\tablenotemark{a}}
  & Saturn  & 5,778  & 5,333  & 28 & 8 & 0.14\%  & 0 & \nodata & \nodata & \nodata & \nodata \\
  & Neptune & 22,335 & 20,095 & 10 & 3 & 0.013\% & 0 & \nodata & \nodata & \nodata & \nodata \\
  & Uranus  & 23,287 & 20,532 & 13 & 2 & 0.0086\% & 0 & \nodata & \nodata & \nodata & \nodata \\
  & Jupiter & 5      & 5      & 0  & 0 & 0\%     & 0 & \nodata & \nodata & \nodata & \nodata \\
\tableline
\shortstack[l]{Scattering TNOs\tablenotemark{a}}
  & Saturn  & \num{4275} & \num{4071} & 90 & 19 & 0.44\% & 0 & \nodata & \nodata & \nodata & \nodata \\
  & Neptune & \num{1476} & \num{1356} & 62 & 18 & 1.22\% & 0 & \nodata & \nodata & \nodata & \nodata \\
  & Uranus  & \num{1304} & \num{1181} & 39 & 2  & 0.15\% & 0 & \nodata & \nodata & \nodata & \nodata \\
\enddata

\tablecomments{Left block: close-encounter sample ($N_\mathrm{cross}$, objects entering one Hill radius). Right block: impactor sample ($N_\mathrm{imp}$). $N_\mathrm{ptg}$ denotes the number of objects falling within at least one LSST pointing. Dots ($\cdots$) mark unavailable entries or planets with no impactors. $N_\mathrm{det}$ is the number of objects detected by LSST (at least observed once) and $N_\mathrm{disc}$ the number linked and discovered.
$\epsilon_\mathrm{disc}=N_\mathrm{disc}/N_\mathrm{input}$ is the discovery efficiency.}
\tablenotetext{a}{Input populations are drawn from the survey-facing catalogs developed in Paper I, excluding unobservable objects to reduce computational cost.}

\end{deluxetable*}
\endgroup

\begin{figure*}[t]
\centering
\includegraphics[width=0.95\linewidth]{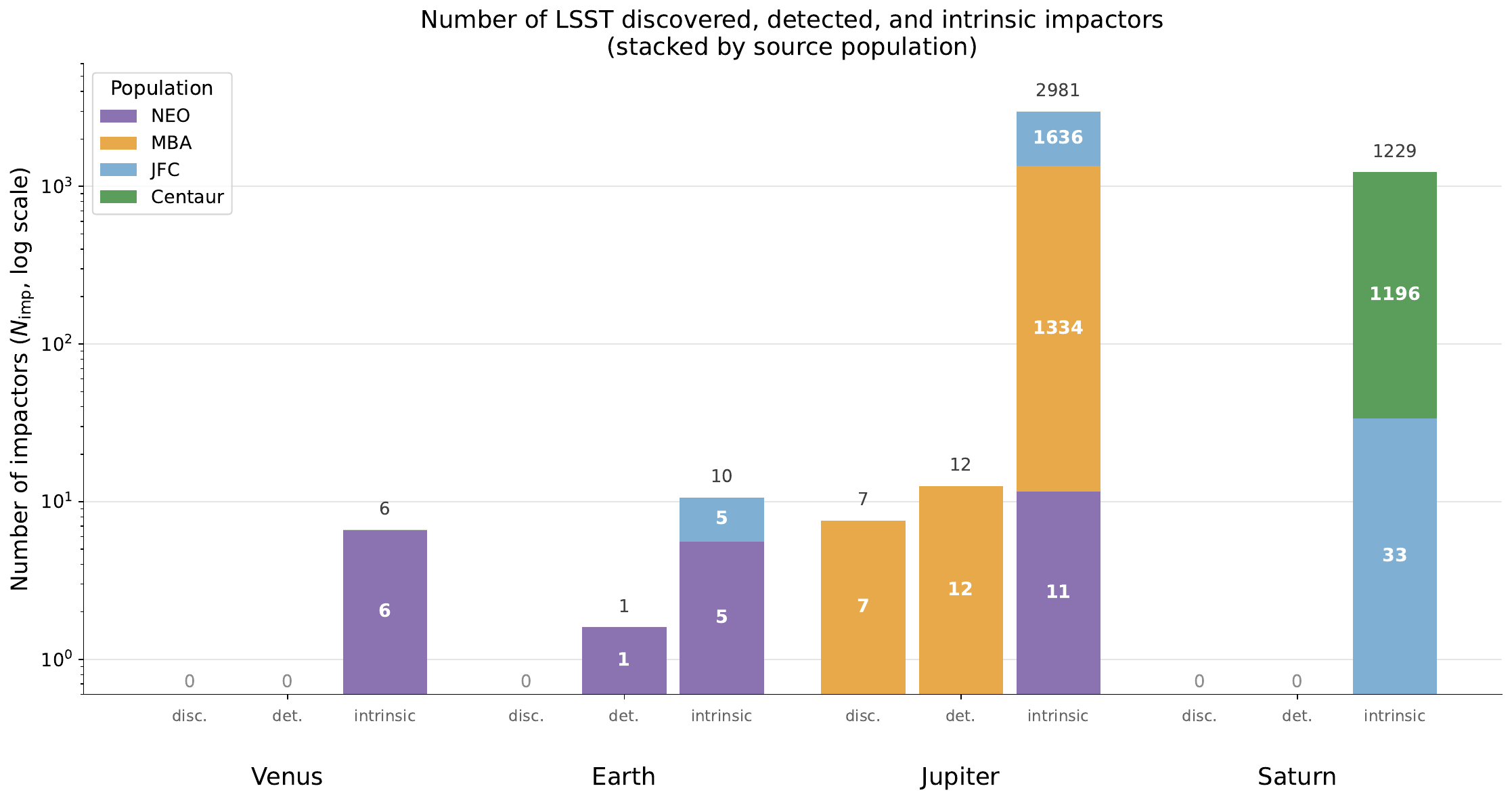}
\caption{Simulated LSST discovered, detected, and intrinsic impactor population, breakdown by source population. Intrinsic populations are taken from Paper I and represent the expected number of $D>10$ m impactors; `det.' denotes objects observed at least once by the LSST simulator, while `disc.' denotes objects that are detected and pass the linking criteria. All counts are based on the periapsis-distance-confirmed sample from Paper I and may be limited by small-number statistics. The uncertainties and corresponding counts derived using the statistical method are presented in Paper I. }
\label{fig:impactorbar}
\end{figure*} 

\subsection{Simulated LSST Discovery Results}
\label{sec:lsst_results}

Table~\ref{tab:loss_modes} presents the discovery efficiencies for close encounters and impactors across the Solar System planets and the Moon, together with the number of objects assigned to each loss mode. Figure~\ref{fig:impactorbar} compares the discovered, detected, and intrinsic populations, with the intrinsic population extrapolated to $D>10$ m. We report detected counts to distinguish objects in LSST observations from those that satisfy the stricter multi-night linking criteria required for the formal discovery. 

Among 59 simulated planetary impactors included in the LSST sample, seven are discovered, all of which are MBA impactors on Jupiter, with diameters ranging from $\sim200$ to $\sim 2000$ m, corresponding to $H_V\simeq 16.24-22.46$. One NEO impactor on Earth is detected but not discovered since it does not satisfy the multi-night linking criterion. The impactor sample is too small to establish a statistically significant population dependence, but it suggests that LSST is more efficient at discovering objects on stable orbits than comet-like or high-eccentricity objects. MBAs, with relatively stable orbits, have higher discovery efficiencies than higher eccentricity populations such as NEOs and JFCs. The latter have shorter observable windows and longer gaps between detectable windows. This pattern follows from LSST’s cadence and moving-object linking requirements, which favor objects that remain observable over multiple visits and nights to form a candidate linkage \citep{Kurlander2025}.

LSST nevertheless discovers a much larger sample of objects whose trajectories undergo planetary close encounters. Approximately $5000$ discovered objects enter Earth's Hill sphere at least once during the 300-year integrations. Of these, 958 undergo a close approach during the 10-year LSST survey interval (2025-2035), and 1146 do so within 20 years (2025-2045). These objects are relevant for planetary defense purposes, since our adopted close-approach distance is five times stricter than the 0.05~au Minimum Orbit Intersection Distance (MOID) threshold used to define Potentially Hazardous Asteroids (PHAs). Among the objects that approach Earth within 10 years, nine exceed the PHA threshold $>140$ m and eight are discovered ($\sim89\%$ discovery rate); 48 are in the $50-140$ m regional-impact-hazard size range and 26 are discovered ($\sim54\%$); 764 are in the $20-50$ m Tunguska-scale hazard size range and 234 are discovered ($\sim31\%$); and 4,111 are in the $10-20$ m Chelyabinsk-scale airburst size range and 687 are discovered ($\sim17\%$). These values are broadly consistent with the LSST discovery efficiencies predicted for the synthetic impactor population at $\sim80\%$, $\sim50\%$, $\sim27\%$, and $\sim11\%$ \citep{Cheng_2026}, respectively. The corresponding encounter samples at the giant planets can similarly constrain the scattering processes that transport trans-Neptunian objects into JFC orbits and, for a small fraction of objects, onward into near-Earth space \citep{VolkMalhotra2008}, thereby clarifying the dynamical precursors of Earth-impacting objects.  

The simulations also predict the discovery of approximately 800 JFCs that undergo planetary encounters, a population-scale sample that complements the recent LSST yield forecasts. Prior work has forecast LSST discovery yields for the NEO, MBA, Jupiter-Trojan, and TNO populations \citep{Kurlander2025} and for the Centaurs \citep{Murtagh2025} but neither study reported a population-level JFC discovery yield. Earlier JFC-specific LSST studies focused primarily on physical characterization, showing that LSST-like sparse photometry can constrain physical properties \citep{Donaldson2024}. Our predicted LSST-discovered JFC sample suggests that such physical characterization could be extended from individual objects to population-level studies of JFCs. 

\begin{figure*}[t]
\centering
\includegraphics[width=0.95\linewidth]{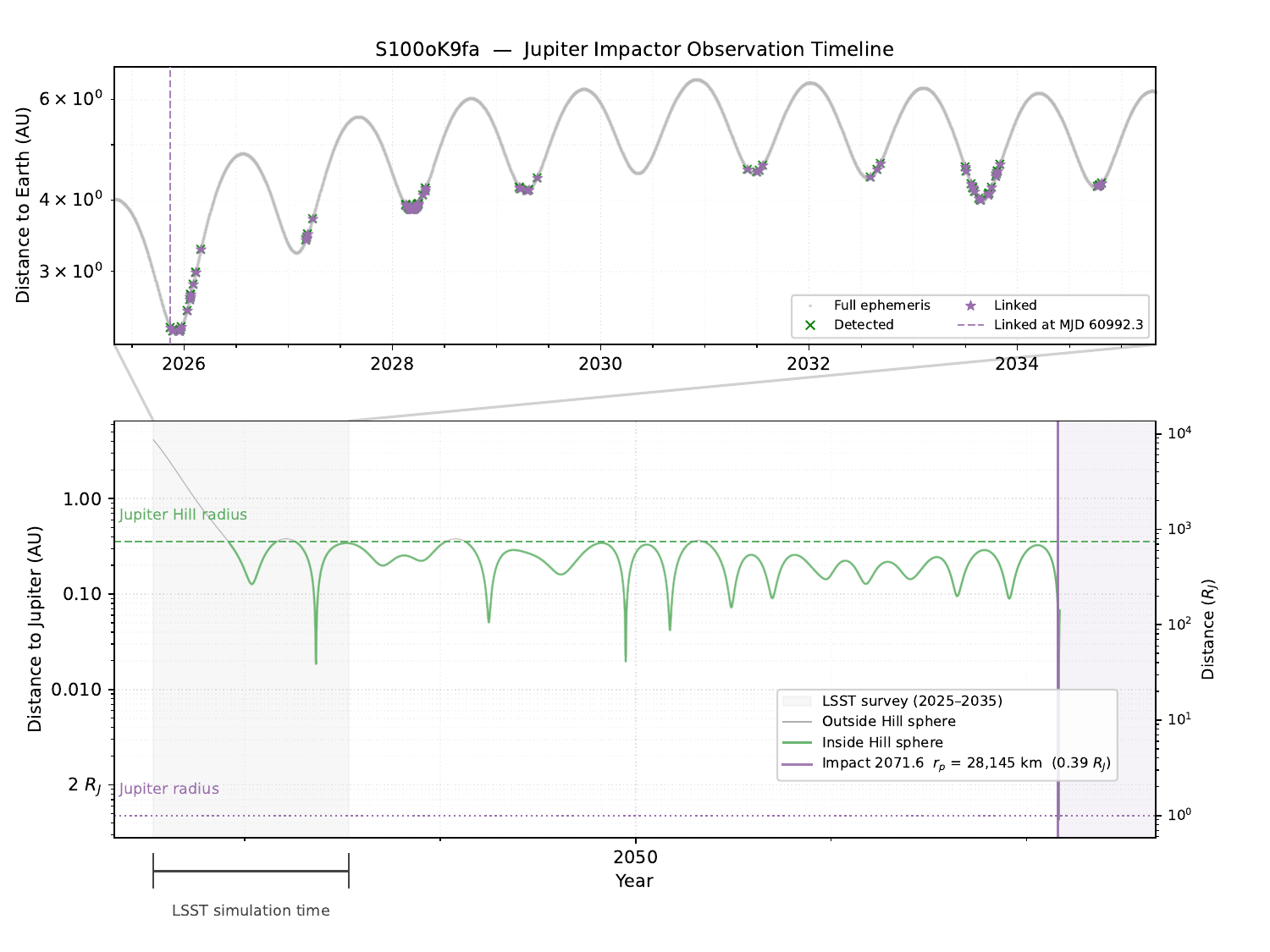}
\caption{Simulated observations for the Jupiter impactor. }
\label{fig:jupiter impactor}
\end{figure*}

\section{Discussion}
\label{sec:discussion}

\subsection{LSST Discovery Forecast with Observation-inferred Impact Rate}

The LSST detection and discovery counts presented in Section~\ref{sec:lsst_results} are based on impactors predicted by models. We denote them as $N_{disc,model}$ and $N_{det,model}$ here. Given the substantial offsets between the model-derived and observation-inferred impact rates, we provide alternative LSST discovery estimates based on the observation-inferred impact frequency, denoted as $N_{disc,obs}$. We derive $N_{disc,obs}$ by multiplying $N_{disc,model}$ or $N_{det,model}$ by the corresponding impact-rate mismatch factor. This scaling assumes that the additional impactors share the same LSST detection or discovery efficiency as the simulated population on average, and is valid only for target bodies where a mismatch has been quantified and $N_{disc,model}$ or $N_{det,model}$ is nonzero, namely Earth and Jupiter.

On Earth, $N_{disc,model}$ is zero, so we substitute $N_{det,model}$ to derive an upper limit for $N_{disc,obs}$, with some caveats. We rescale the single detected count by the Earth impact-rate mismatch factor of $2.2-21$ and obtain an observation-inferred detection of $\sim2-21$ impactors larger than $10$ m. We treat this as an upper limit on $N_{disc,obs}$, because discovered objects form a subset of detected objects. As an independent comparison based on real CNEOS-recorded impactors, \citet{Chow_2026} predict that LSST will discover $1-2$ meter-sized and larger ($\gtrsim 1$~m) imminent impactors, broadly consistent the with higher end of our prediction. 

For Jupiter, $N_{disc,model}$ is seven and we scale it by the impact-rate mismatch factor of $2.6-47.7$ to predict $N_{disc,obs}$ $\sim18-334$. LSST does not report a detection count because only the identified objects will pass the alert system, so we report the detected count only as a reference. Applying the same scaling to $N_{det,model}$ of 12, we predict $N_{det,obs}$ $\sim31-572$. 



We caution that, for planets that have zero $N_{det,model}$, for instance, Mars and the Moon, the true $N_{disc,obs}$ or $N_{det,obs}$ could exceed zero with the higher observation-inferred impact frequency, but our scaling method cannot recover these cases.

We predict that the discovery-to-collision warning time widens on both ends relative to the prediction in Section~\ref{sec:lsst_results}, because the observation-inferred impactors could span the full $>10$ m range, and warning time scales with size \citep{Cheng_2026}. We also expect a shorter warning time on average, because the size frequency skews towards the smaller-size end. 

\subsection{LSST Offers Opportunities to Observe Jupiter Impactors before They Collide}

Figure~\ref{fig:jupiter impactor} gives an example of a bound Jupiter impactor discovered by LSST before collision, based on the \texttt{Sorcha} simulation results in Section \ref{sec:lsst_results}. The simulated observations marked in the plot demonstrate the possibility of studying the intricate bound dynamics of Jupiter impactors using LSST. Such observations are particularly valuable because it is often challenging to reconstruct the bound or unbound state after an impact. For SL9, more than 112,000 backward integrations produced nearly equal probabilities of a direct heliocentric impact and a prior jovicentric capture, 47\% and 53\%, respectively \citep{SanchezLavega2010}. Although LSST and \texttt{Sorcha} are not designed to monitor short-duration Jovian impact flashes, both due to cadence and the brightest magnitude limit cut, they can detect faint progenitors before the terminal impacts. The seven discovered Jupiter impactors in our simulation have discovery-to-impact lead times of $1.94-257.4$ yr within the 300-yr impact integrations, assuming no cometary activity, and individual objects receive as many as $\sim400$ detections during the 10-yr LSST survey. Two of the seven objects are detected by LSST while traversing Jupiter's Hill sphere, providing rare opportunities to observe the transition between heliocentric and jovicentric motion directly. Detailed case studies of these objects could constrain the capture duration and transition from temporary capture to collision, which cannot be recovered from an impact flash or atmospheric scar alone.

\subsection{Limitations of Unmodeled Cometary Effects}

Our forward model treats every object as an inert body with a fixed absolute magnitude and a purely gravitational trajectory. This assumption is well suited to the asteroidal populations, but it neglects cometary activity, which is most relevant for the volatile-rich source populations such as JFCs, Centaurs, and scattering TNOs. We summarize below the principal effects we do not model and, where possible, the direction in which each biases our results.  

First, we do not model activity-driven brightening. Our \texttt{Sorcha} setting takes each object as a point source scaled from its assigned absolute magnitude, so an active object surrounded by a coma is far brighter and easier to detect than the bare nucleus we assume, and our discovery efficiencies for the JFC and Centaur populations are therefore conservative in this respect.   

Second, we do not model mass loss or fragmentation. Comets lose mass through sublimation, can split or disrupt entirely, and become dormant on timescales shorter than their dynamical lifetimes. The scale of the mass loss can be large at decameter sizes. The observed comet 450P/LONEOS with an effective nucleus radius $R_{\rm N}=1.8\pm0.5$~km has a dust-loss rate of $\sim4-8~{\rm kg\,s^{-1}}$ \citep{Schambeau2026}. Applying the same absolute rate to a 10-m body implies $\Delta R\sim2-4$~m over 300~yr for $\rho\simeq500~{\rm kg\,m^{-3}}$. A continuously active 10-m nucleus could thus lose $\sim80$--$98\%$ of its mass over our integration window, and even at $10\%$ activity, the nucleus would lose $\sim10$--$20\%$ of its mass. An object counted as a $D>10$~m impactor in our model may therefore shrink its size or cease to exist before it collides. Our modeled impactor and encounter counts for the volatile-rich populations can be overestimated. 


Third, we do not include the non-gravitational accelerations produced by outgassing. Non-gravitational accelerations are small per orbit but accumulate over the multi-century integrations, and the same encounter-driven comet-activity activation identified by \citet{Lilly2024} would strengthen their activity for objects about to impact. They can therefore alter which objects approach a planet or impact, and shift the discovery-to-impact lead times we report for the Jupiter impactors.

These effects act in competing directions. Cometary activity-driven brightening raises detectability while mass loss and disruption reduce the surviving population. A dedicated model coupling an activity and mass-loss model to the dynamical integrations is a natural extension of this work. 

\section{Conclusions}
To answer whether known objects yield a plausible impact count and whether LSST offers pre-impact observing opportunities, we ran known small-body populations from the JPL SBDB through the  Paper~I impact-search pipeline, and we forward modeled the synthetic planetary close-encounter and impactor populations from Paper~I through the LSST survey using \texttt{Sorcha}. Our main findings are as follows.

Our pipeline recovers 13 confirmed impact events, of which 12 are single-body Earth impactors and one multi-fragment Jupiter impactor. It reproduces the most recent sub-meter Earth impactor \texttt{2026~JN4}, with a predicted impact time within $10.3$ minutes of the observed May 2026 atmospheric entry over the Arafura Sea, and the other 11 past Earth impactors before 2026. It also recovers 19 of the 21 known Shoemaker Levy 9 fragments, each within $\sim20-26$~minutes of its published 1994 Jupiter impact time. Our pipeline also yields two predictions that cannot yet be verified. \texttt{2015~FK488}, a $\sim1.8$~km Centaur, is predicted to strike Jupiter in 2227 and is the only predicted impactor above 10~m; the NEO impactor \texttt{2022~KG1} is predicted to reach Earth in 2266. Both objects have poorly constrained orbits propagated over century timescales, and thus require full impact-probability propagation to verify the impact and to assess the need for follow-up observations.
 
Assuming modeled impact rates, we predict that LSST will discover seven synthetic planetary impactors, and all seven are main-belt asteroids that strike Jupiter, with diameters of $\sim200$--$2000$~m ($H_V\simeq16.2$--$22.5$). One Earth-impacting NEO is detected but fails the multi-night linking criterion. The seven discovered synthetic Jupiter impactors have discovery-to-impact lead times of $1.94-257.4$~yr, and two are detected while traversing Jupiter's Hill sphere. These objects offer rare opportunities to observe the transition from temporary capture to collision directly, a state that is difficult to reconstruct from an impact flash or atmospheric scar alone. 

If we instead assume the observation-inferred impact rates and scale the discovery counts with the impact-rate mismatch factor, LSST will discover at most $\sim20$ Earth impactors and $\sim300$ Jupiter impactors larger than 10 m. We also expect wider warning time windows given the larger size range of these impactors. 

LSST discovers impactors on stable orbits more efficiently than high-eccentricity and comet-like objects. This pattern follows the LSST cadence and its moving-object linking requirements, which favor objects that remain observable across multiple visits and nights.

Additionally, the simulations predict $\sim800$ discovered Jupiter-family comets that undergo planetary encounters, a sample size that could extend physical characterization from individual objects to the JFC population as a whole. 


\begin{acknowledgments}
Q.C. and D.S. acknowledge support from the Duke University Trinity College of Arts and Sciences Department of Physics and from the Cosmology Group. D.S. acknowledges support from the Duke University Electrical and Computer Engineering Department.

D.S. is supported by the Department of Energy grant DE-SC0010007, the David and Lucile Packard Foundation, the Templeton Foundation, and Sloan Foundation.
\end{acknowledgments}


\bibliography{ref}{}
\bibliographystyle{aasjournalv7.1}

\end{document}